\documentclass[a4paper,11pt]{article}
\usepackage{amsmath,amssymb,amsfonts,amsthm}%
\usepackage{csquotes}
\usepackage{multirow}
\usepackage{float}
\usepackage{graphicx}
\usepackage{bbm}
\usepackage{diagbox}
\usepackage{mathtools}
\usepackage{array}
\usepackage{stmaryrd}
\usepackage{scrlayer-scrpage}
\usepackage{xcolor}
\usepackage{geometry}
\usepackage{authblk}
\usepackage{hyperref} 
\usepackage{mathrsfs}%
\usepackage[title]{appendix}%
\usepackage{xcolor}%
\usepackage{textcomp}%
\usepackage{manyfoot}%
\usepackage{booktabs}%
\usepackage{listings}%
\usepackage{url}%
\begin{document}

\title{Effective dimensionality reduction for Greeks computation using Randomized QMC}

\author[1, \dag]{Luca~Albieri}
\author[2]{Sergei~Kucherenko}
\author[3]{Stefano~Scoleri}
\author[4,5,\dag]{Marco~Bianchetti}

\affil[1]{Risk Management, DXT Commodities SA, Lugano, Switzerland \vspace{1mm}}
\affil[2]{Imperial College London, London, SW72AZ, UK\vspace{1mm}}
\affil[3]{Intesa Sanpaolo, IMI Corporate and Investment Banking, Equity, Commodities and FX models \vspace{1mm}}
\affil[4]{Financial \& Market Risk Management, Intesa Sanpaolo, Milan, Italy \vspace{1mm}}
\affil[5]{Department of Statistical Sciences \enquote{Paolo Fortunati}, University of Bologna, Italy \vspace{1mm}}

\affil[$\dag$]{Corresponding author, \texttt{marco.bianchetti@unibo.it}, \texttt{lalbieri33@gmail.com}}

\date{}
\maketitle
\begin{abstract}
Global sensitivity analysis is employed to evaluate the effective dimension reduction achieved through Chebyshev interpolation and the conditional pathwise method for Greek estimation of discretely monitored barrier options and arithmetic average Asian options. We compare results from finite difference and Monte Carlo methods with those obtained by using randomized Quasi Monte Carlo combined with Brownian bridge discretization. Additionally, we investigate the benefits of incorporating importance sampling with either the finite difference or Chebyshev interpolation methods. Our findings demonstrate that the reduced effective dimensionality identified through global sensitivity analysis explains the performance advantages of one approach over another. Specifically, the increased smoothness provided by Chebyshev or conditional pathwise methods enhances the convergence rate of randomized Quasi Monte Carlo integration, leading to the significant increase of accuracy and reduced computational costs. \\

\vspace{1cm}

\end{abstract}
\vspace{0.5cm}

\newpage
\tableofcontents

\newpage
\section{Introduction}
In modern finance, risk management depends significantly on advanced market and counterparty risk measures obtained through multi-dimensional, multi-step Monte Carlo simulations. These measures are crucial for risk management practices, particularly as they are frequently required for regulatory compliance and are subject to stringent validation by regulatory authorities. Producing daily price and risk assessments for large portfolios with numerous counterparties requires substantial computational resources and the adoption of complex frameworks and industrial-grade methodologies.\\

\noindent
The Monte Carlo (MC) method is commonly applied in pricing complex financial instruments when analytical solutions are not feasible. MC simulation provides a practical and flexible approach by randomly sampling input space parameters of the payoff function and approximating the instrument's fair value as the average of the simulated outcomes. However, MC simulation is often time-consuming and has a slow convergence rate, as indicated by the decay of the root-mean-square error $\mathcal{O}(1/\sqrt{N})$, where $N$ is the number of simulated paths. While variance reduction techniques have been developed to enhance simulation efficiency and decrease root-mean-square error, they do not accelerate the convergence rate (\cite{Gla03}, \cite{Jac01}).\\

\noindent
The Quasi Monte Carlo (QMC) method employs low discrepancy sequences instead of pseudo-random numbers to sample the input variable space more uniformly. Numerous studies have demonstrated the superior effectiveness of QMC with Sobol'sequences (\cite{Sob11}) compared to standard Monte Carlo methods (\cite{Bia15}, \cite{Ren20}). QMC theoretically ensures a convergence rate greater or equal to Monte Carlo, when combined with Brownian Bridge discretization or other effective dimension reduction techniques (e.g., PCA), it shows superior performance over simple Monte Carlo (\cite{Caf97}, \cite{Bia15}, \cite{Sco21}). Despite its advantages, the QMC method lacks an in-sample error estimation. The randomized QMC (RQMC) overcomes this obstacle by providing a practical error bound by randomly scrambling the low discrepancy sequence. In \cite{Hok22} RQMC was combined with the Brownian bridge and a local volatility model to estimate the price and delta of a geometric Asian option, demonstrating a marked improvement over standard QMC methods. \cite{Gir07} demonstrated the effectiveness of combining RQMC with importance sampling and conditional expectation techniques for variance reduction, with this hybrid approach producing the most accurate results in barrier option pricing. We focus primarily on financial sensitivities, adopting a similar approach while refining dimension reduction techniques to further enhance convergence. It was demonstrated in many papers (\textit{e.g.} \cite{Hok22}) that Owen's scrambling (\cite{Owe97}) yields the highest efficiency among other randomisation techniques and for this reason it was used in  our implementation.\\

\noindent
In the context of estimating financial sensitivities through Monte Carlo simulations, two main approaches are commonly used (\cite{Gla03}): methods that require re-pricing over a grid of values for the differentiation parameter (e.g., finite differences) and those that avoid re-pricing (e.g., pathwise and likelihood ratio methods). We explored two techniques, one from each category, and compared them with the standard finite difference method.
The first technique relies on Chebyshev interpolation (CI), introduced in \cite{Mar21}. This method estimates Greeks (risk sensitivities) using a grid of Chebyshev points and has been applied across a variety of exotic financial instruments. The approach also includes an adaptive algorithm for defining an appropriate interpolation domain. CI offers computational efficiency and produces smoother Greeks. However, optimizing the grid to strike a balance between bias and variance is essential, as it is with finite differences.
The second technique is the conditional pathwise (CPW) method, described in \cite{Zha20}. This method smooths the payoff function via a conditioning step, allowing the interchange of expectation and differentiation operators. It has been shown to be effective in the Black-Scholes framework for exotic payoffs like Asian, Barrier, and Lookback options. QMC integration benefits significantly from smoother payoff functions, improving convergence in most cases (\cite{Zha20}, \cite{Bil22}). Although highly effective, conditional pathwise may be limited by certain model dynamics and regularity assumptions.
In our research, we applied these methods to estimate financial sensitivities, focusing on using dimension-reduction techniques to enhance convergence.\\

\noindent
Global Sensitivity Analysis (GSA) is a highly effective tool for investigating complex models, offering a comprehensive method for model analysis. Unlike traditional sensitivity analysis, which focuses on local behaviour by calculating the derivative of a function $f$ at a specific point $x^*$ in the domain, GSA provides a broader perspective. It explores the entire domain without the need to specify any particular point, making it a global analysis. Additionally, GSA evaluates the effect of varying one input (or a combination of inputs) while simultaneously accounting for variations in other inputs, allowing for the measurement of interactions between variables (\cite{Sob93}, \cite{Sob01}). 
This methodology pinpoints pivotal parameters whose uncertainty mostly affects the output and provides insights into variable interactions. Such insights facilitate variable prioritization, elimination of non-essential variables, and reduction of problem effective dimensions. GSA is based on sensitivity indices called Sobol' indices which are divided into two types: main effect indices, which gauge the individual contribution of each input parameter to output variance, and total sensitivity indices, which quantify the collective contribution of either a single input factor or a group of inputs. Sobol' indices enable the definition of dimensional indicators for the problem under exam, such as effective dimension and average dimension. This approach highlights how effective dimensional reduction is linked to convergence improvements in QMC integration (\cite{Kuc07}, \cite{Bia15}).\\

\noindent
The objectives of our work are the following:
\begin{itemize}
    \item[1)] To determine whether the performance improvement of CI over finite difference methods, as observed in \cite{Mar21} could be further enhanced by randomized QMC integration;
    \item[2)] Assess possible effective dimension reductions offered by CI and CPW methods. 
\end{itemize}

\noindent
These research questions are driven by two key factors: firstly, QMC and randomized QMC integration have shown higher convergence rates when the target function is sufficiently smooth (\cite{Owe97}); secondly, when the effective dimension is significantly reduced, QMC and randomized QMC outperform traditional Monte Carlo methods (\cite{Kuc07}, \cite{Bia15}). 

\noindent
This paper is structured as follows: Chapter 2 introduces the methodologies used in our study, focusing on Monte Carlo simulation techniques and the estimation of financial sensitivities. Chapter 3 discusses the GSA method and provides formulas for computing Sobol' indices. Chapter 4 presents numerical results on the estimation of Sobol' indices and convergence analysis for Asian and Barrier options. Chapter 5 concludes by summarizing the findings of the study. Finally, Appendix A offers a concise summary of the conditional estimates for Down-and-Out Call and Asian Call options. Appendix B details the parameters for Greeks approximation. Appendix C discusses the theory of Importance Sampling for Down-and-Out Call options, while Appendix D presents the results for Prices.

\section{Theoretical Framework}
We introduce some key notations. Let $(\Omega, \mathcal{F}, \mathbb{Q})$ represent the probability space with the risk-neutral measure $\mathbb{Q}$ and filtration $\mathcal{F}=\{\mathcal{F}_t\}_{t\in [0,T]}$. Here, $T$ stands for the maturity date, and $\theta$ represents the set of pricing and model parameters. The underlying asset process is denoted by $\mathbf{S}_t = \{S_t\}_{t\in [0,T]}$, and $r_t$ refers to the interest rate used for discounting. We express the price of the financial instrument as $V_t(\theta)$:

\begin{equation*}
\begin{split}
    &V_t(\theta) = \mathbb{E}^{\mathbb{Q}}[D(t,T)\mathcal{P}(\mathbf{S}_t, \theta)|\mathcal{F}_t] = \mathbb{E}^{\mathbb{Q}}_t[D(t,T)\mathcal{P}(\mathbf{S}_t, \theta)].\\
    &D(t,T) = exp\bigg(-\int_t^T r_tdt\bigg).
\end{split}
\end{equation*}
Here, $D(t,T)$ represents the discount factor between times $t$ and $T$, while $\mathcal{P}(S_t, \theta)$ denotes the payoff function of the instrument. Within the context of the Black-Scholes model, we discretize the dynamics of the underlying asset process as follows:

\begin{equation}
\label{Euler_BS}
    S_{j+1} = S_j \exp\bigg[\bigg(r -\frac{\sigma^2}{2}\bigg) \Delta t_j + \sigma \Delta W_j \bigg].
\end{equation}
The volatility and interest rate are kept constant and the time grid $\{t_j\}_{j=1,..,D}$ is equispaced with fixed step $\Delta t_j \equiv \Delta t$.
Concerning the Wiener process discretization, two schemes are compared: Euler and Brownian Bridge schemes. The Brownian Bridge discretization plays a crucial role in exploiting the improved uniformity properties of LDS and scrambled LDS in the definition of the most important steps in the Wiener process.
\begin{table}[H]
\centering
\renewcommand{\arraystretch}{2}
\begin{tabular}{|p{7.5cm}|p{7.5cm}|}
\hline
\textbf{Euler} & \textbf{Brownian Bridge} \\
\hline
$\Delta W_j = \sqrt{\Delta t} \xi_j,  \quad\quad \xi_j \sim N(0,1)$ & $W_N=\sqrt{\Delta t_{N0}}\xi_1, \quad\quad \xi_j \sim N(0,1)$  \\
$W_j = W_{j-1} + \Delta W_j$ & $W_j=\frac{\Delta t_{kj}}{\Delta t_{ki}}W_i + \frac{\Delta t_{ji}}{\Delta t_{ki}}W_k + \sqrt{\frac{\Delta t_{kj}\Delta t_{ji}}{\Delta t_{ki}}}\xi_l$\\
\hline
\end{tabular}
\caption{Wiener process discretization schemes. For Brownian Bridge scheme $\Delta t_{ij} = t_i - t_j$.}
\label{tab:discretizations}
\end{table}

\subsection{Monte Carlo}
We compared two simulation approaches for generating multidimensional uniform samples. In a standard MC approach, we construct a $D \times N$ matrix of uniformly distributed variables $\mathbf{U}$, where $D$ represents the number of simulated risk factors and $N$ is the number of scenarios. The uniform sample $\mathbf{U}$ is then transformed into the desired distribution (e.g., normal) using the inverse cumulative distribution function: $\mathbf{X} = \Phi^{-1}(\mathbf{U})$. Finally, we estimate the payoff for each simulated scenario and estimate the desired integral as the average of all the outcomes:

\begin{equation}
    V_{N}^{MC}(\theta) = \frac{1}{N} \sum\limits_{i=1}^{N} f(\mathbf{U}_i, \theta) 
\end{equation}
The error bounds in a standard MC simulation are $\varepsilon_{N}^{MC} = \sigma_f/\sqrt{N}$ with $\sigma_f$ denoting the standard deviation of the discounted payoff $f(\mathbf{U}, \theta)$. A Mersenne Twister pseudo-random number generator was employed to generate the uniform samples $\mathbf{U}$.

\subsection{Randomized QMC}
The RQMC method enhances error estimation by scrambling low-discrepancy sequence (LDS) points, similar to the MC approach. We employed the Owen algorithm of scrambling Sobol' LDSs provided by BRODA Ltd. (\cite{BRODA},  \cite{Ata21}). Owen's scrambling has been identified as optimal for LDS scrambling in financial applications (see \textit{e.g.} \cite{Hok22}. For an LDS $\mathbf{\Tilde{U}}$ of dimensions $D \times n$, we apply Owen's scrambling $K$ times to obtain $K$ scrambled LDSs $\{\mathbf{\Tilde{Q}}_k\}_{k=1,...,K}$. Subsequently, we compute $K$ distinct estimates of the instrument's price and take the average of these results:

\begin{equation}
     V_{k,n}^{RQMC}(\theta) = \frac{1}{n} \sum\limits_{i=1}^{n} f(\mathbf{Q}_{k,i}, \theta), \quad\quad V_{n}^{RQMC}(\theta) = \frac{1}{K} \sum\limits_{i=1}^{K} V_{i,n}^{RQMC}(\theta).
\end{equation}
The error estimation for RQMC is then defined as follows:
\begin{equation}
\label{error_RQMC}
    \varepsilon_{n,K}^{RQMC}=\frac{\sigma_f}{\sqrt{K}}, \quad\quad \sigma_{f}^2 = \frac{1}{K-1} \sum\limits_{k=1}^{K}\bigg(V_{n}^{RQMC} - V_{k,n}^{RQMC}\bigg)^2.
\end{equation}
To compare the error estimation between MC and RQMC methods, we ensure that $nK = N$.

\subsection{Financial Greeks estimation}
Sensitivities to risk factors or "Greeks" are derivatives of the instrument's fair value w.r.t. specific parameters. Greeks are primarily employed for hedging purposes in financial risk management. We focus on the following set of Greeks:
\begin{equation}
    \begin{split}
        &Delta = \frac{\partial V}{ \partial S_0}, \quad\quad Gamma = \frac{\partial^2 V}{ \partial S_0^2},\\
        &Vega = \frac{\partial V}{ \partial \sigma}, \quad\quad Vomma = \frac{\partial^2 V}{ \partial \sigma^2}.
    \end{split}
\end{equation}
Techniques for calculating financial Greeks generally fall into two main categories \cite{Gla03}: those that involve repricing over a grid of nodes for the differentiation parameter, and those that do not. In the first category, managing the bias-variance trade-off is crucial. This entails optimizing the parameters to balance the bias from the derivative approximation and the variance caused by the simulation. While increasing the number of nodes reduces bias, it also raises computational costs. The finite difference (FD) method, a widely used and straightforward technique, belongs to this category. Detailed discussions on the FD method can be found in \cite{Gla03} and \cite{Jac01}. Given a shift parameter $h > 0$, the 2-point and 3-point central finite difference approximations for first and second-order derivatives are defined as:
\begin{equation}
\label{Fd_central}
    \begin{split}
        \frac{\partial V}{\partial \theta} &\simeq \frac{V(\theta + h) - V(\theta - h)}{2h},\\
        \frac{\partial^2 V}{\partial \theta^2} &\simeq \frac{V(\theta + h) - 2V(\theta) + V(\theta - h)}{h^2}.
    \end{split}
\end{equation}
\noindent
An alternative approach to grid selection was proposed in \cite{Mar21}, where Chebyshev points were used instead of a uniformly spaced grid. In \cite{Tre04}, a more general framework for arbitrary grids with $L$ points is presented. Building on the chosen grid, one can utilize a recursive definition for the differential matrix of order $m$, denoted by $D^m$ (\cite{Wel97}, \cite{Tre04}), which leads to the sensitivity estimation: 

\begin{equation}
    \frac{\partial^m V}{\partial \theta^m} \simeq D^m \overline{V}_N^{\mathcal{X}}, \quad\quad \overline{V}_N^{\mathcal{X}} = (V_{1,N}^{\mathcal{X}}, ..., V_{L,N}^{\mathcal{X}}).
\end{equation}
Here $V_{i, N}^{\mathcal{X}}$ the i-th grid point estimation with $N$ scenarios adopting simulation method ${\mathcal{X}}$ (MC or RQMC). We note that this approach applies to any grid and the computation of matrix $D^m$ only depends on the chosen grid. \\

\noindent
The second category of methods for computing financial Greeks includes the so-called \textit{pathwise} (PW) method (\cite{Gla03}, \cite{Jac01}). This technique is based on interchanging the differentiation and expectation operators:

\begin{equation}
\label{switch}
    \frac{\partial}{\partial \theta} \mathbb{E}^{\mathbb{Q}}_t[f(\mathbf{U}, \theta)] = \mathbb{E}^{\mathbb{Q}}_t\bigg[ \frac{\partial}{\partial \theta}  f(\mathbf{U}, \theta)\bigg].
\end{equation}

\noindent
Typically, equation (\ref{switch}) is valid under certain regularity conditions (discussed and formulated in \cite{Gla03}). However, the standard PW method is not suitable for cases with discontinuous payoff functions, such as barrier options. A comprehensive review of research advancements on pathwise and likelihood ratio approaches for financial sensitivities is provided in \cite{Zha20}. Additionally, the authors introduced the \textit{conditional pathwise} (CPW) method, which includes a conditioning step that smooths the payoff function before applying the derivative and mean operators. We denote $\mathbf{U}_{-i} = (U_1, ..., U_{i-1}, U_{i+1}, ..., U_{D})$.

\begin{equation}
    \label{conditioning_step}
    G(\mathbf{U}_{-i}, \theta) = \mathbb{E}_t[f(\mathbf{U}, \theta)| \mathbf{U}_{-i}] , \quad\quad \mathbb{E}_t[f(\mathbf{U}, \theta)] = \mathbb{E}_t[G(\mathbf{U}_{-i}, \theta)].
\end{equation}
The conditioning step (\ref{conditioning_step}) requires a \textit{separating variable condition} to hold, which decomposes the payoff function as follows:
\begin{equation}
\label{var_sep_cond}
    f(\mathbf{X}, \theta) = h(\mathbf{X}, \theta) \mathbbm{1}_{\{\psi_d(\mathbf{X}_{-i}, \theta) < X_i < \psi_u(\mathbf{X}_{-i}, \theta)\}}
\end{equation}
for suitable functions $h$, $\psi_d$, and $\psi_u$ (see Appendix A for a detailed example). In this section, we express the discounted payoff function in terms of the normal random variable $\mathbf{X} = \Phi^{-1}(\mathbf{U})$, rather than using the uniform random variable $\mathbf{U}$ as in the previous sections. In \cite{Zha20}, it was demonstrated that under the Black-Scholes model, many payoffs satisfy the condition (\ref{var_sep_cond}). When simulating the underlying process $S_t$ within the CPW framework, equation (\ref{Euler_BS}) must be modified as follows:
\begin{equation}
\begin{split}
       \Tilde{S}_j &= S_0 \exp{\big(\mu(t_j-t_1) + \sigma (W_j - W_1)\big)} \\
       &= S_j \exp{\big(-\mu t_1 - \sigma W_1\big)}\\
       &= S_j \exp{\big(-\mu t_1 - \sigma X_1 \sqrt{t_1}\big)}.
\end{split}
\end{equation}
Here $\mu = r - \frac{\sigma^2}{2}$.
With this new formulation the process $\Tilde{S}_j$ will depend only on random variables $X_2,..., X_d$ allowing to perform a variable separation. 
Additional regularity conditions are necessary for the CPW method to be applied; Theorem 3.1 in \cite{Zha20} provides a detailed statement of these conditions. This theorem specifies requirements on the payoff function $f(\mathbf{X}, \theta)$ and the functions $\psi_d$ and $\psi_u$ to ensure the differentiability of $G(\mathbf{X}_{-i}, \theta)$, in (\ref{conditioning_step}). Appendix A provides explicit CPW Greek formulas for selected payoff functions. When using QMC or RQMC simulation, the convergence behaviour of the CPW method is significantly improved if the target function is sufficiently regular.

\section{Global sensitivity analysis}
Many practical problems involve functions with highly complex structures. GSA offers insights into the overall structure of a function by quantifying how variations in input variables affect the output variables (\cite{Sob01}, \cite{Kuc07}). GSA plays a crucial role in predicting and explaining the benefits observed with QMC and Randomized QMC integration over standard MC methods. GSA is based on ANOVA decomposition, which, for a square-integrable function $g: [0,1]^D \longrightarrow \mathbb{R}$, provides an expansion for the variance $\sigma_{g}^2$ as a sum of partial variances due to each subset of variables:
\begin{equation}
    \label{ANOVA}
    \sigma_{g}^2 = \sum_{\mathbf{u}\subseteq \mathbf{D}} \sigma_{\mathbf{u}}^2.
\end{equation}
We denote a subset of indices as $\mathbf{u} \subseteq \{1,..,D\} = \mathbf{D}$ with cardinality $\#\mathbf{u}=u$ and $\overline{\mathbf{u}}$ its complementary set of variables. 
The ANOVA decomposition is a highly effective method for analyzing functions with multiple variables, particularly when determining which variables have the greatest impact on the function's output. Sobol' indices, also known as global sensitivity indices, are defined as the following ratios:
\begin{equation}
\label{Sobol_indices}
    S_{\mathbf{u}} = \frac{\sigma_{\mathbf{u}}^2}{\sigma_g^2},\quad \quad
    S_{\mathbf{u}}^{tot} = 1-S_{\overline{\mathbf{u}}}.
\end{equation}
These indices are used to rank variables based on their influence on a function $g(\cdot)$, allowing unimportant variables to be fixed and higher-order terms in the decomposition (\ref{ANOVA}) to be discarded. The main effect and total-order Sobol' indices provide sufficient information to assess whether a variable is significant or negligible  (Table \ref{tab:model_dependence}).
\begin{table}[H]
\centering
\renewcommand{\arraystretch}{1.5}
\begin{tabular}{|p{2cm}|p{4.5cm}|}
\hline
\textbf{Case} & \textbf{Variable Impact} \\
\hline
$S_{i}^{tot}\simeq0$ & Negligible effect of $x_i$. \\\hline
$S_{i}\simeq1$ & Dominant effect of $x_i$.\\\hline
$S_{i}\simeq S_{i}^{tot}$ & Low interaction of $x_i$ with other variables.\\\hline
\end{tabular}
\caption{Cases of model dependence that arise in practical experiments.}
\label{tab:model_dependence}
\end{table}

\subsection{Effective dimension}
The concept of effective dimensions was introduced in \cite{Caf97}. These indicators provide insights into the influence of input variables in a complex mathematical model, regardless of its nominal dimension. When dealing with financial instrument pricing and sensitivity estimation, it was observed that a lower effective dimension ensures that QMC or RQMC integration outperforms the standard MC (\cite{Caf97}, \cite{Bia15},\cite{Ren20}). The effective dimension in the \textit{superposition sense} is defined as the smallest integer $d_S$ such that
\begin{equation}
    \label{super_eff_dim}
    \sum\limits_{\# \mathbf{u} < d_S} S_{\mathbf{u}} \geq 1 - \varepsilon
\end{equation}
for some threshold $\varepsilon > 0$ and does not depend on the variable's sampling order. The effective dimension in the \textit{truncation sense} is defined as the smallest integer $d_T$ such that
\begin{equation}
    \label{trunc_eff_dim}
    \sum\limits_{\mathbf{u} \subseteq \{1,2,...,d_T\}} S_{\mathbf{u}} \geq 1 - \varepsilon.
\end{equation}
Sobol' indices can be used to categorize functions into three groups, as summarized in Table \ref{tab:function_types} (\cite{Kuc11}). For functions of type A and B, QMC outperforms MC even in high-dimensional cases. However, for functions of type C, QMC loses its advantage over MC due to the significance of higher-order terms in the corresponding ANOVA decomposition.\\

\noindent
A direct calculation of the effective dimensions in both the superposition and truncation senses based on their definitions is impractical. A useful alternative effective dimensionality metric, which can take fractional values, is the \textit{average dimension} $d_A$ (\cite{Owe03}), defined as:  
\begin{equation}
    d_A = \sum\limits_{\mathbf{u} \subseteq \mathbf{D}} \# \mathbf{u} S_{\mathbf{u}}
\end{equation}
which could be computed directly from total effect indices as
\begin{equation}
    d_A = \sum\limits_{i=1}^{D} S_{i}^{tot}.
\end{equation}
In this study, we employed the \textit{average dimension} $d_A$ as an indicator of effective dimension, due to its simpler computation. We note, that the reduction concerns the effective dimension, rather than the nominal dimension (as commonly seen in $e.g.$ standard PCA applications where only the most influential combinations of variables are selected). In our case, we do not reduce the nominal dimension of simulated variables. Instead, we assess the impact of variable subsets using Sobol' indices and map the initial dimensions of Sobol' LDS points which are better distributed than higher dimensions to the most important variables in BBD or PCA sampling schemes.

\begin{table}[H]
\centering
\renewcommand{\arraystretch}{2}
\begin{tabular}{|p{1cm}|p{4cm}|p{4cm}| p{4cm}|}
\hline
\textbf{Type} & \textbf{Property} & \textbf{Eff. Dimension} & \textbf{Interaction} \\
\hline
\textbf{A} & $S_{\mathbf{u}}^{tot}/\# \mathbf{u} \gg S_{\overline{\mathbf{u}}}^{tot}/\# \overline{\mathbf{u}}$ & $d_S \leq d_T \ll D$ & \footnotesize Few important variables and dominant low order interactions.\\\hline
\textbf{B} & $S_i\simeq S_i^{tot},\quad \sum\limits_{i=1}^{D} S_i \simeq 1$ & $d_S \ll D,  d_T \simeq D$ & \footnotesize Dominant low-order
interactions and equally important variables.\\\hline
\textbf{C} & $S_i\ll S_i^{tot},\quad \sum\limits_{i=1}^{D} S_i \ll 1$ & $d_S \simeq d_T \simeq D$ & \footnotesize Dominant high-order
interactions and equally important variables.\\
\hline
\end{tabular}
\caption{Function classification according to dependence on variables importance and their interactions.}
\label{tab:function_types}
\end{table}

\subsection{Computing Sobol' indices}
The main and total effect indices can be computed by MC or QMC approximation of D-dimensional integrals (\cite{Sob01}). Here we assume that function $g$ is defined in the m-dimensional unit hypercube $[0,1]^m$. Let $\mathbf{v} = \{u_{i_1}, ..., u_{i_d} \} \subseteq \mathbf{u}$ with $i_j \in \mathbf{D}$ for all $j$ and $\mathbf{w}$ such that $(\mathbf{v}, \mathbf{w}) = \mathbf{u}$. Samples $\mathbf{u}=(\mathbf{v}, \mathbf{w})$ and $\mathbf{u}^{\prime}=(\mathbf{v}^{\prime}, \mathbf{w}^{\prime})$ are assumed to be independent. We adopted total effect indices formulas from \cite{Sob01} and main effect indices with improved formulas introduced in \cite{Mau07}: 
\begin{equation}
    \label{sobol_indices}
    \begin{split}
    S_{\mathbf{u}}^{tot} &= \frac{1}{2 \sigma^2_g} \int_0^1 \big[g(\mathbf{u}) - g(\mathbf{v}^{\prime}, \mathbf{w})\big]^2 d\mathbf{u}d\mathbf{v}^{\prime} \simeq \frac{1}{2 \sigma^2_g} \bigg(\frac{1}{N} \sum\limits_{j=1}^{N} \big[ g(\mathbf{u}_j) - g(\mathbf{v}_j^{\prime}, \mathbf{w}) \big]^2\bigg),\\
        S_{\mathbf{u}} &= \frac{1}{\sigma^2_g} \int_0^1 g(\mathbf{u})\big[g(\mathbf{v}, \mathbf{w}^{\prime}) - g(\mathbf{u}^{\prime})\big] d\mathbf{u}^{\prime}d\mathbf{u} \simeq \frac{1}{\sigma^2_g} \bigg(\frac{1}{N} \sum\limits_{j=1}^{N} g(\mathbf{u}_j)\big[ g(\mathbf{v}_j, \mathbf{w}_j^{\prime}) - g(\mathbf{u}_j^{\prime}) \big]\bigg).
    \end{split} 
\end{equation}
Here $\sigma^2_g$ denotes the function variance. 


\section{Numerical results}
We tested the introduced methodologies on two different options: an arithmetic Asian Call $\mathcal{C}_{a}$ and a discretely monitored Down-Out Call barrier option $\mathcal{C}_{do}$ with the following payoffs: 
\begin{equation}
    \begin{split}
        \mathcal{C}_{a} &= \max\big(\overline{S}-K, 0\big), \quad\quad \overline{S} = \frac{1}{D}\sum\limits_{j=1}^D S_j,\\
        \mathcal{C}_{do} &= \max\big(S_D-K, 0\big)\mathbbm{1}_{\min S_j > B}. 
    \end{split}
\end{equation}
The pricing parameters used for the two payoffs are provided in Table \ref{tab:pricing_parameters}.

\begin{table}[H]
    \centering
\begin{tabular}{| c | c | c | c | c |}
\hline
 \textbf{Variable} & \textbf{Symbol} & \textbf{Asian Call} & \textbf{Down-Out Call} \\
 \hline
 Maturity & $T$ & $0.25y$  & $0.25y$\\
  \hline
 Spot Price & $S_0$ & $100$ & $100$\\
  \hline
Time steps & $D$ & 32 & 32\\
  \hline 
Interest rate & $r$ & $3\%$ & $3\%$\\
  \hline 
Volatility & $\sigma$ & $30\%$ & $30\%$ \\
  \hline 
Strike & $K$ & 100 & 100 \\
  \hline 
Barrier & $B$ & - & 90/80/70\\
  \hline 
\end{tabular}
\caption{Pricing parameters for the financial instruments under analysis.}
\label{tab:pricing_parameters}
\end{table}

\noindent
The simulation methods employed were MC and RQMC, as outlined in Section 2. Error convergence was assessed through in-sample error estimations for both methods, eliminating the need for analytical or near-exact benchmarks. The Greeks considered include first and second-order derivatives with respect to spot price and volatility (Delta, Gamma, Vega, and Vomma).

\noindent
We compared three approaches for estimating Greeks. The first one is the central finite difference method, which uses 2 points for first-order derivatives and 3 points for second-order derivatives. The second approach is CI, utilizing a 7-point grid (refer to \cite{Mar21} for details on this method). The third method is the conditional pathwise (CPW) approach introduced in Section 3. Additionally, we combined both finite difference and CI with importance sampling for the barrier option. Importance sampling (IS), a variance reduction technique, smooths the payoff function and mitigates the impact of singularities on convergence (\cite{Gla03}, \cite{Zha20}). Additional details on the importance sampling adopted for Barrier Options is presented in Appendix C. To evaluate the convergence behavior and singularity effects on barrier option Greeks, we used three different barrier levels for the Down-Out Call. For completeness, we also report prices estimation and errors for Asian and Barrier Options in Appendix D.

\subsection{Effective dimension}
We begin by presenting the GSA metrics for the Greeks of both payoffs, calculated using QMC integration with $2^{18} = 262144$ points, as per formulas (\ref{sobol_indices}). The indices are computed specifically for the barrier option with a barrier level of $B=90$ and the Asian option with a strike price of $K=100$. Our observations reveal that the barrier option is highly sensitive to the barrier level, leading us to focus on dimensional reduction for the nearest spot price and barrier levels. This analysis helps us identify the most effective methods for this challenging parameter choice. Conversely, the Asian option's error behavior is not notably influenced by the strike level. Based on the average dimension $d_A$, we can infer the potential benefits of RQMC over standard MC. Empirical evidence suggests that RQMC tends to outperform MC when $d_A<3$, a finding we confirm in our study, noting that the smoothness of the payoff function also significantly impacts convergence. In the following table, for each Greek, we report results for ten different combinations of discretization algorithm (ED or BBD), differentiation approach (FD, CI or CPW) and addition of importance sampling (IS).

\begin{table}[H]
\centering  
\begin{tabular}{| c  c | c | c | c | c |}
\hline
\multicolumn{2}{|c|}{\textbf{Down-Out Call}} & $\mathbf{S_i/S_i^{tot}}$ & $\mathbf{\sum S_i^{}}$ & $\mathbf{d_A}$ & \textbf{Type}\\
\hline
\textbf{Delta} & \textbf{ED + FD} & $10^{-4}\shortrightarrow .04$ & .07 & 11.29 & C\\
 & \textbf{BBD + FD} & $10^{-4}\shortrightarrow 1$ & .1 & 5.05 & C\\
 & \textbf{ED + CI} & $10^{-3}\shortrightarrow .16$ & .28 & 7.04 & C\\
 & \textbf{BBD + CI} & $10^{-4}\shortrightarrow 1$ & .4 & 3.1 &  C\\
  & \textbf{ED + CPW} & $10^{-2}\shortrightarrow 1$ & .66 &  2.53 & B\\
 & \textbf{BBD + CPW} & $10^{-7}\shortrightarrow 1$ & .8 & 1.43 & A\\
 & \textbf{ED + FD + IS} & $.27\shortrightarrow 0.31$  &  .76 & 2.61 & B\\
 & \textbf{BBD + FD + IS} & $10^{-3}\shortrightarrow 1$ & .98 & 1.01 & A\\
 & \textbf{ED + CI + IS} & $.28\shortrightarrow 0.32$ & .77 & 2.53 & B\\
 & \textbf{BBD + CI + IS} & $10^{-4}\shortrightarrow 1$ & .98 & 1.01 & A\\
 \hline
 \textbf{Gamma} & \textbf{ED + FD} & $10^{-5}\shortrightarrow 10^{-2}$ & .02 & 12.56 & C\\
 & \textbf{BBD + FD} & $10^{-4}\shortrightarrow 1$ & .02 & 5.68 & C\\
 & \textbf{ED + CI}  & $10^{-4}\shortrightarrow 10^{-2}$ & .04 & 10.97 & C\\
 & \textbf{BBD + CI} & $10^{-4}\shortrightarrow 1$ & .06 & 4.40 & C\\
  & \textbf{ED + CPW} & $10^{-4}\shortrightarrow 1$ & .03 & 7.57 & C\\
 & \textbf{BBD + CPW} & $10^{-7}\shortrightarrow .62$ & .23 & 3.23 & C\\
  & \textbf{ED + FD + IS} & $10^{-4}\shortrightarrow 10^{-3}$ & .06 & 16 & C\\
 & \textbf{BBD + FD + IS} & $10^{-3}\shortrightarrow 1$ & .81& 1.29 & A\\
 & \textbf{ED + CI + IS} & $10^{-5}\shortrightarrow 10^{-3}$ & .06 & 16 & C\\
 & \textbf{BBD + CI + IS} & $10^{-3}\shortrightarrow 1$ & .82 & 1.26 & A\\
 \hline
 \textbf{Vega} & \textbf{ED + FD} & $10^{-4}\shortrightarrow 10^{-3}$ & .03 & 12.27 & C\\
 & \textbf{BBD + FD} & $10^{-12}\shortrightarrow 1$ & .03 & 5.9 & C\\
 & \textbf{ED + CI} & $10^{-2}\shortrightarrow .12$ & .33 & 4.55 & C\\
 & \textbf{BBD + CI} & $10^{-3}\shortrightarrow 1$ & .57 & 2.37 & C\\
  & \textbf{ED + CPW} & $.19\shortrightarrow 1$ & .47 & 2.11 & B\\
 & \textbf{BBD + CPW} & $10^{-9}\shortrightarrow 1$ & .79 & 1.38 & A\\
  & \textbf{ED + FD + IS} & $.41\shortrightarrow .49$ & .65 & 1.45 & B\\
 & \textbf{BBD + FD + IS} & $10^{-3}\shortrightarrow 1$ & 1 & 1 & A\\
 & \textbf{ED + CI + IS} & $.41\shortrightarrow .49$ & .65 & 1.45 & B\\
 & \textbf{BBD + CI + IS} & $10^{-3}\shortrightarrow 1$ & 1 & 1 & A\\
 \hline
 \textbf{Vomma} & \textbf{ED + FD} & $10^{-4}\shortrightarrow 10^{-3}$ & .02 & 13.55 & C\\
 & \textbf{BBD + FD} & $10^{-10}\shortrightarrow 1$ & .01 & 5.69 & C\\
 & \textbf{ED + CI} & $10^{-4}\shortrightarrow 10^{-2}$ & .03 & 10.87 & C\\
 & \textbf{BBD + CI} & $10^{-5}\shortrightarrow .22$ & .03 & 4.8 & C\\
  & \textbf{ED + CPW} & $10^{-4}\shortrightarrow 1$ & .07 & 6.69 & C\\
 & \textbf{BBD + CPW} & $10^{-9}\shortrightarrow .23$ & .14 & 3.52 & C\\
  & \textbf{ED + FD + IS} & $.01\shortrightarrow .06$ & .13 & 2.97 & B\\
 & \textbf{BBD + FD + IS} & $10^{-3}\shortrightarrow 1$ & .99 & 1.03 & A\\
 & \textbf{ED + CI + IS} & $.01\shortrightarrow .05$ & .12 & 2.9 & B\\
 & \textbf{BBD + CI + IS} &$10^{-3}\shortrightarrow 1$ & .98 & 1.03 & A\\
 \hline
\end{tabular}
\caption{Main to total effect index ratios: arrow "$ \rightarrow$" in the column for $S_i / S_i^{\text {tot}}$ denotes the change in the value with the increase of index $i$; 
sum of main indices; average dimension and estimated function types as defined in Table \ref{tab:function_types} for Down-Out Call Greeks using ten different methodologies. Here ED stands for Euler discretization.}
\label{tab:gsa_do_call}
\end{table}
\noindent
Table (\ref{tab:gsa_do_call}) presents GSA indicators for Down-and-Out Call option. BBD leads to a significant dimensional reduction compared to ED. CI appears to further decrease the average dimension $d_A$ relative to the finite difference method. The greatest dimensional reductions are observed with CPW, particularly when combined with BBD. First-order derivatives display lower average dimensions and reduced variable interactions, whereas second-order derivatives exhibit higher effective dimensionality and more notable higher-order interactions. In all cases, importance sampling plays a crucial role in achieving effective dimensional reduction.

\begin{table}[H]
\centering  
\begin{tabular}{| c  c | c | c | c | c |}
\hline
\multicolumn{2}{|c|}{\textbf{Asian Call}} & $\mathbf{S_i/S_i^{tot}}$ & $\mathbf{\sum S_i^{}}$ & $\mathbf{d_A}$ & \textbf{Type}\\
\hline
\textbf{Delta} & \textbf{ED + FD} & $10^{-3}\shortrightarrow .25$ & .66 & 3.88 & B\\
 & \textbf{BBD + FD} & $10^{-4}\shortrightarrow .69$ & .75 & 1.71 & A\\
 & \textbf{ED + CI} & $10^{-2}\shortrightarrow .35$ & .75 & 2.45 & B \\
 & \textbf{BBD + CI} & $10^{-2}\shortrightarrow 1$ & .8 & 1.31 & A \\
  & \textbf{ED + CPW} & $10^{-2}\shortrightarrow 1$ & .83 & 1.6 & B \\
 & \textbf{BBD + CPW} & $.35\shortrightarrow .84$ & .88 & 1.18 & A \\
 \hline
 \textbf{Gamma} & \textbf{ED + FD} & $10^{-5}\shortrightarrow 10^{-3}$ & .05 & 30.37 & C \\
 & \textbf{BBD + FD} & $10^{-5}\shortrightarrow 10^{-3}$ & .05 & 21.41 & C \\
 & \textbf{ED + CI} & $10^{-4}\shortrightarrow 10^{-2}$ & .04 & 12.97 & C\\
 & \textbf{BBD + CI} & $10^{-4}\shortrightarrow .23$ & .11 & 3.17 & C\\
  & \textbf{ED + CPW} & $10^{-4}\shortrightarrow 1$ & .05 & 5.87 & C\\
 & \textbf{BBD + CPW} & $10^{-4}\shortrightarrow 1$ & .26 & 2.34 & C\\\hline
 \textbf{Vega} & \textbf{ED + FD} & $.26\shortrightarrow .51$ & .69 & 1.42 & B\\
 & \textbf{BBD + FD} & $10^{-2}\shortrightarrow 1$ & .87 & 1.14 & A\\
 & \textbf{ED + CI} & $.39\shortrightarrow .51$ & .69 & 1.42 & B\\
 & \textbf{BBD + CI} & $10^{-2}\shortrightarrow 1$ & .87 & 1.14 & A\\
  & \textbf{ED + CPW} & $.41\shortrightarrow 1$ & .71 & 1.34 & A\\
 & \textbf{BBD + CPW} & $.12\shortrightarrow 1$ & .87 & 1.13 & A\\\hline
 \textbf{Vomma} & \textbf{ED + FD} & $10^{-5}\shortrightarrow .03$ & .07 & 27.2 & C\\
 & \textbf{BBD + FD} & $10^{-4}\shortrightarrow 10^{-2}$ & .13 & 27.32 & C\\
 & \textbf{ED + CI} & $10^{-3}\shortrightarrow 10^{-2}$ & .15 & 4.89 & C\\
 & \textbf{BBD + CI} & $10^{-4}\shortrightarrow .61$ & .63 & 2.52 & C\\
  & \textbf{ED + CPW} & $10^{-2}\shortrightarrow 1$ & .18 & 2.23 & C\\
 & \textbf{BBD + CPW} & $10^{-3}\shortrightarrow .67$ & .67 & 1.37 & B\\
 \hline
\end{tabular}
\caption{Main to total effect index ratios: arrow "$ \rightarrow$" in the column for $S_i / S_i^{\text {tot}}$ denotes the change in the value with the increase of index $i$; sum of main indices; average dimension and estimated function types as defined in Table \ref{tab:function_types} for Asian Call Greeks using six different methodologies.}
\label{tab:gsa_asian_call}
\end{table}
\noindent
Table \ref{tab:gsa_asian_call} presents GSA indicators for Greeks of the Asian Call option.
The results for the Asian option indicate a low average dimension for the first-order derivatives, with BBD typically providing the best performance. For the second-order derivatives, CI offers a significant effective dimensional reduction compared to the finite difference method, yielding particularly strong results when combined with BBD. CPW combined with BBD provides the lowest average dimensions.

\newpage
\subsection{Convergence Analysis}
In this Section, we report the results and error analysis of the two simulation methods (MC with Euler discretization ({MC+ED}) and randomized QMC with Brownian Bridge discretization ({RQMC+BBD}) combined with four Greeks estimation approaches (finite difference FD, Chebyshev interpolation CI and conditional pathwise CPW). 
Plots present log errors vs. the number of simulations taken at $N = 2^p$ for $p = 10, ..., 18$ and linear regression lines derived from log errors.
These regressions are based on the assumption that the error follows a law of the form (for some constant $c$):
\begin{equation}
    \varepsilon_N = \frac{c}{N^\alpha}, \quad\quad \log_{10}(\varepsilon_N) = \log_{10}(c) - \alpha \log_{10}(N).
\end{equation}
\noindent
Lines' slopes ($\mathbf{\alpha}$) are an estimate for the convergence rates of each methodology. 
Constant $c$ defines the intercepts of regression lines and also provide useful information about the efficiency of the QMC (RQMC) and MC methods: lower intercepts mean that the simulated value
starts closer to the exact value. We also report the error ($\mathbf{\varepsilon_0}$) for $2^{10}$ simulations to quantify the starting error magnitude. The parameters adopted for FD and CHEB are discussed in Appendix B.

\subsubsection{Down-Out Call Option}
\begin{figure}[H]
\centering
\includegraphics[scale=0.33]{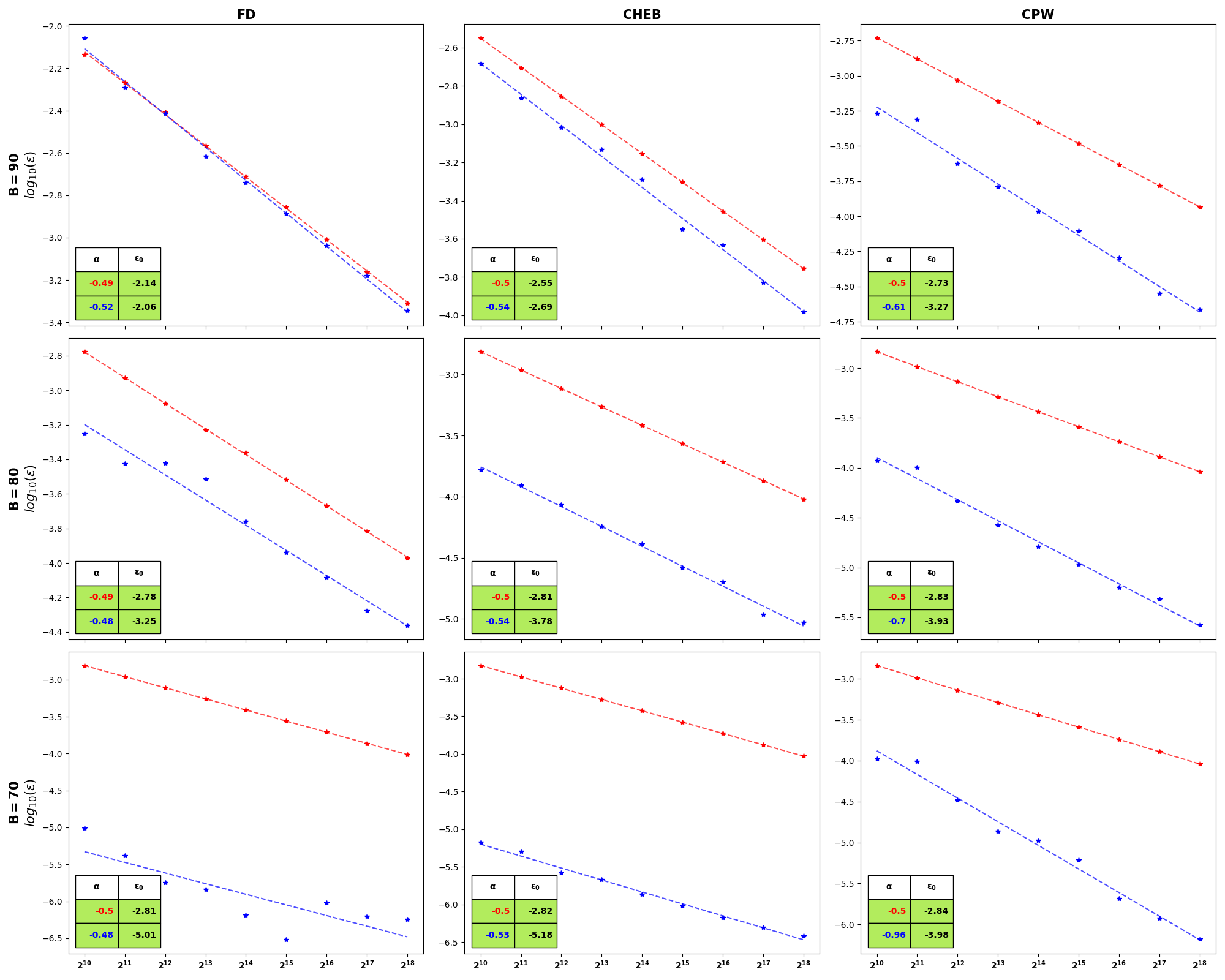}
\caption{Convergence rates for Down-Out Call barrier option Delta with barrier levels at 90, 80 and 70 adopting \textbf{MC+ED} (red) and \textbf{RQMC+BBD} (blue).}
\label{fig:delta_conv_SingleBarrier}
\end{figure}

\begin{figure}[H]
\centering
\includegraphics[scale=0.33]{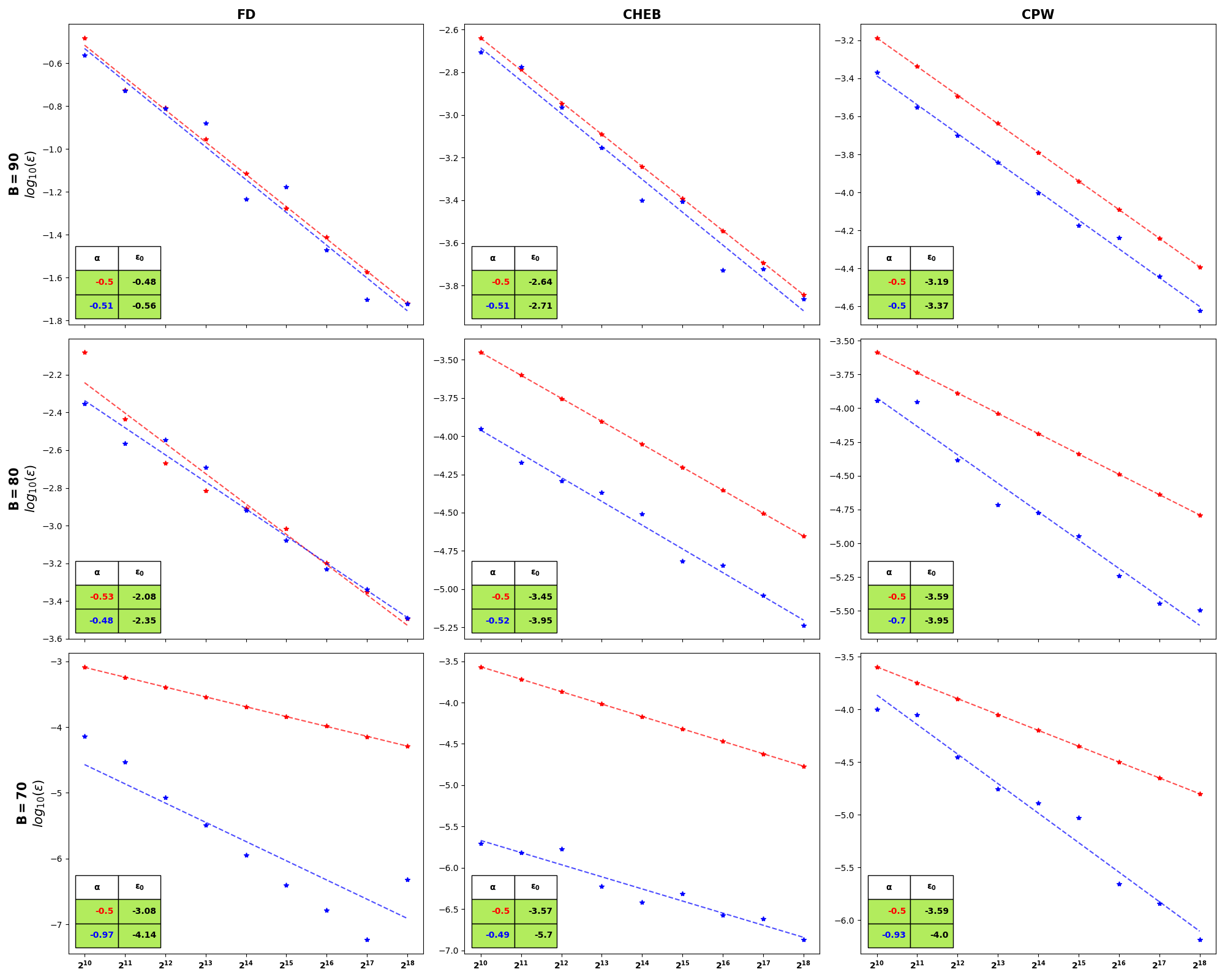}
\caption{Convergence rates for Down-Out Call barrier option Gamma with barrier levels at 90, 80 and 70 adopting \textbf{MC+ED} (red) and \textbf{RQMC+BBD} (blue).}
\label{fig:gamma_conv_SingleBarrier}
\end{figure}

\begin{figure}[H]
\centering
\includegraphics[scale=0.33]{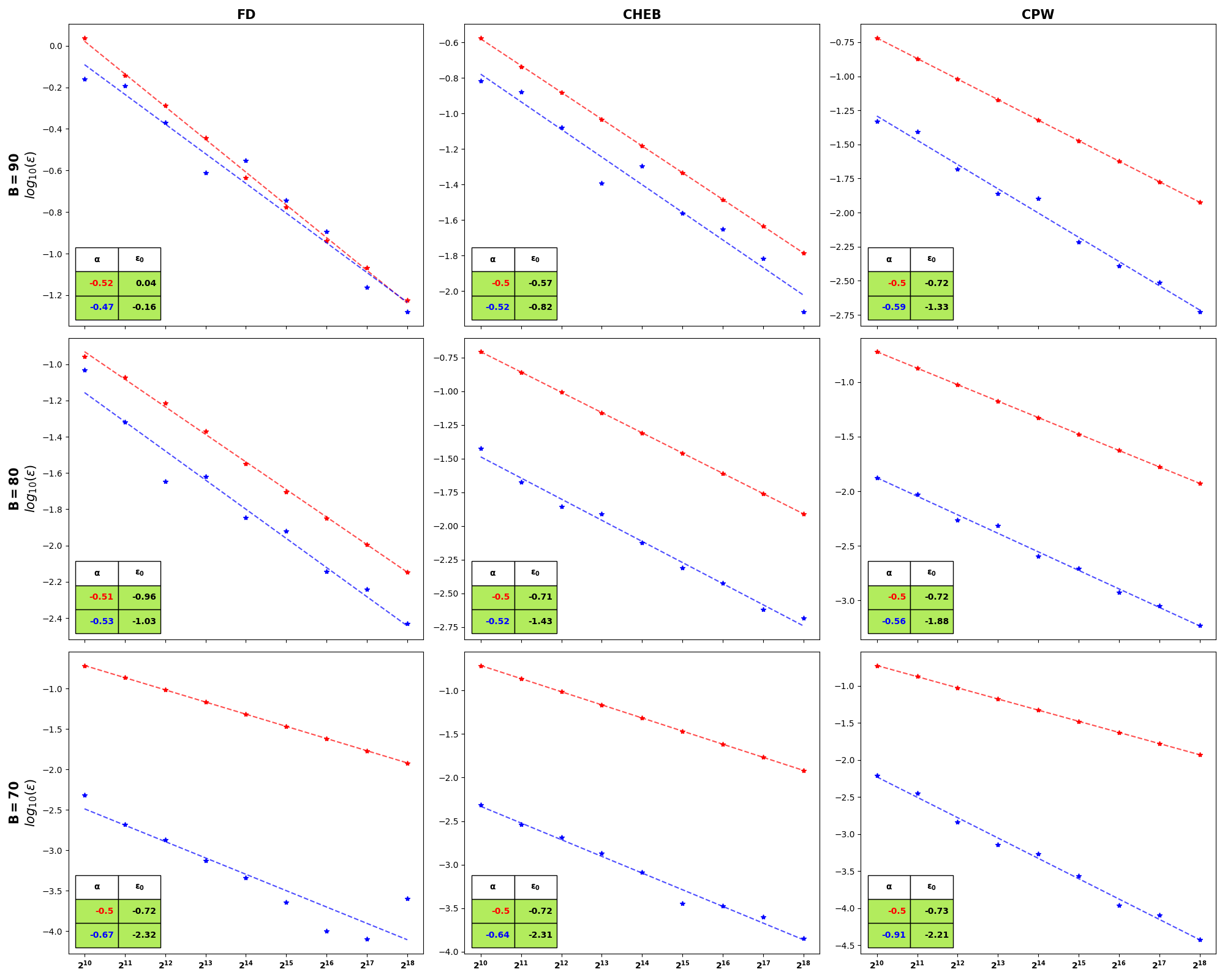}
\caption{Convergence rates for Down-Out Call barrier option Vega with barrier levels at 90, 80 and 70 adopting \textbf{MC+ED} (red) and \textbf{RQMC+BBD} (blue).}
\label{fig:vega_conv_SingleBarrier}
\end{figure}

\begin{figure}[H]
\centering
\includegraphics[scale=0.33]{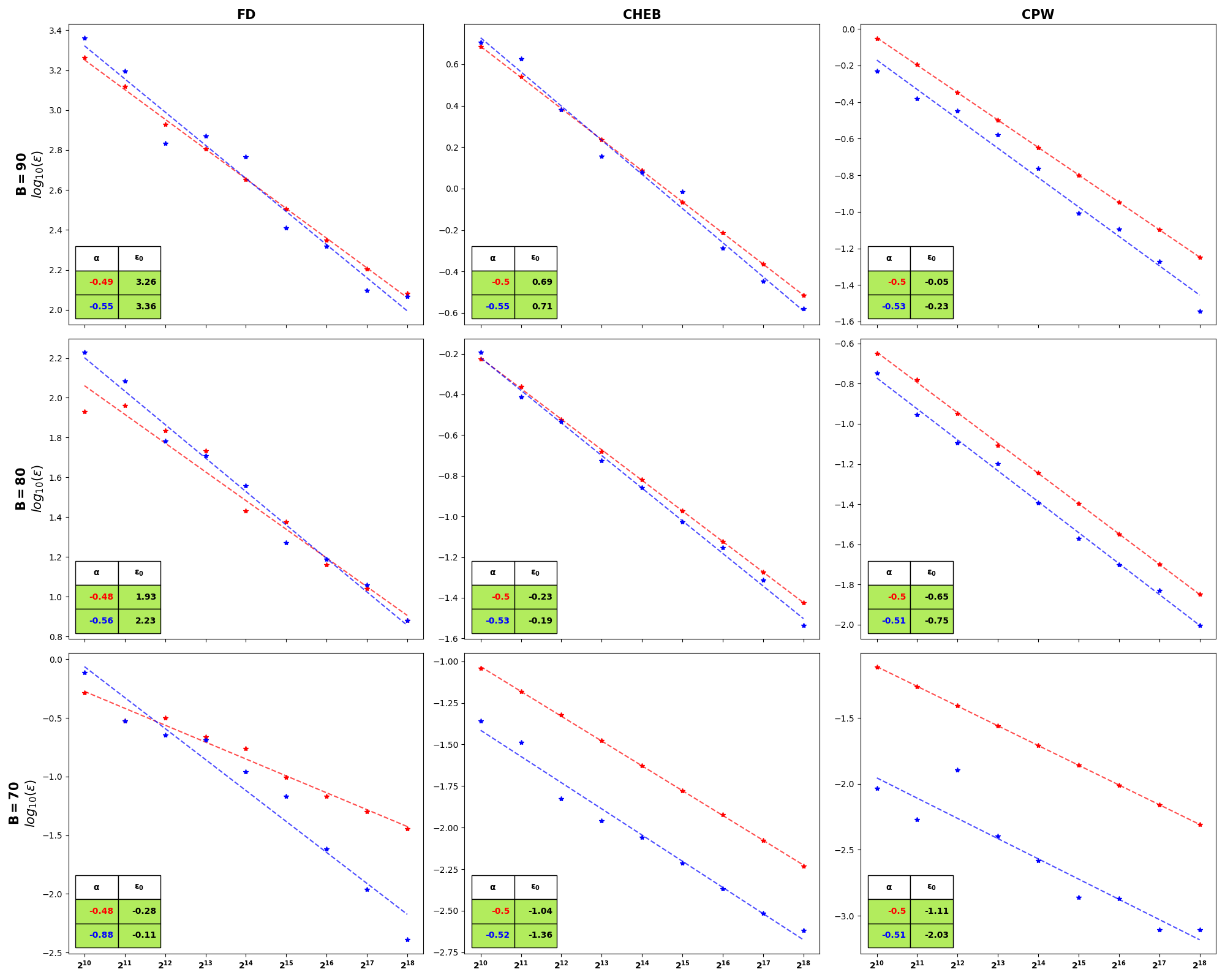}
\caption{Convergence rates for Down-Out Call barrier option Vomma with barrier levels at 90, 80 and 70 adopting \textbf{MC+ED} (red) and \textbf{RQMC+BBD} (blue).}
\label{fig:vomma_conv_SingleBarrier}
\end{figure}

Figures \ref{fig:delta_conv_SingleBarrier}-\ref{fig:vomma_conv_SingleBarrier} contain the linear regression lines for the log error of the Down-Out Call option. As expected, the convergence rates for MC+ED are approximately 0.5 in all cases. In each case, the greater the difference between the barrier level $B$ and the spot price $S_0$, the better the convergence. However, for second-order derivatives, the estimated errors exhibit noise, which affects the reliability of the slopes. In almost all cases, RQMC+BBD significantly outperforms MC+ED. The exceptions occur when the barrier level is closer to the spot price; in these instances, the reduced smoothness of the payoff limits the improvement, with the most significant benefit coming from selecting the appropriate Greek estimation method, which reduces the initial error but does not improve the convergence rate. The convergence is generally faster when adopting CPW combined with RQMC and BBD. The advantages of CI over FD with a 3-point grid are reduced when considering a 7-point equispaced grid but, in most cases, CI offers a lower starting error.

\subsubsection{Arithmetic Asian Call Option}

For the Asian option, we report the convergence rates and initial errors for the four Greeks considering a strike value of 100\footnote{Various strike values were tested, showing a negligible effect on the convergence behavior; therefore, these results are not included here}. We observe an improved accuracy in adopting CI compared to FD for second-order Greeks, with an increased convergence rate in the case of Delta and Gamma. Remarkably, for Gamma and Vomma, the CI and CPW error with just $2^{10}$ simulations is comparable to the FD error produced with $2^{18}$ simulations. Generally, CPW leads to the highest convergence rates except for Gamma, where CI produces the best results. The enhanced smoothness and reduced effective dimension provided by CI and CPW result in higher convergence rates when using RQMC in most cases.

\subsubsection{Importance sampling}

We combined importance sampling with FD and CI for the Down-Out Call barrier option with a barrier value of 90. The adoption of importance sampling further enhances the smoothness and reduces the average dimension leading to both reduced starting errors and increased convergence rates for RQMC+BBD for all the Greeks. The difference in performance between FD and CI is eliminated, which removes the need for more nodes in the Greek approximation and decrease computational efforts.

\newpage
\noindent

\begin{figure}[H]
\centering
\includegraphics[scale=0.45]{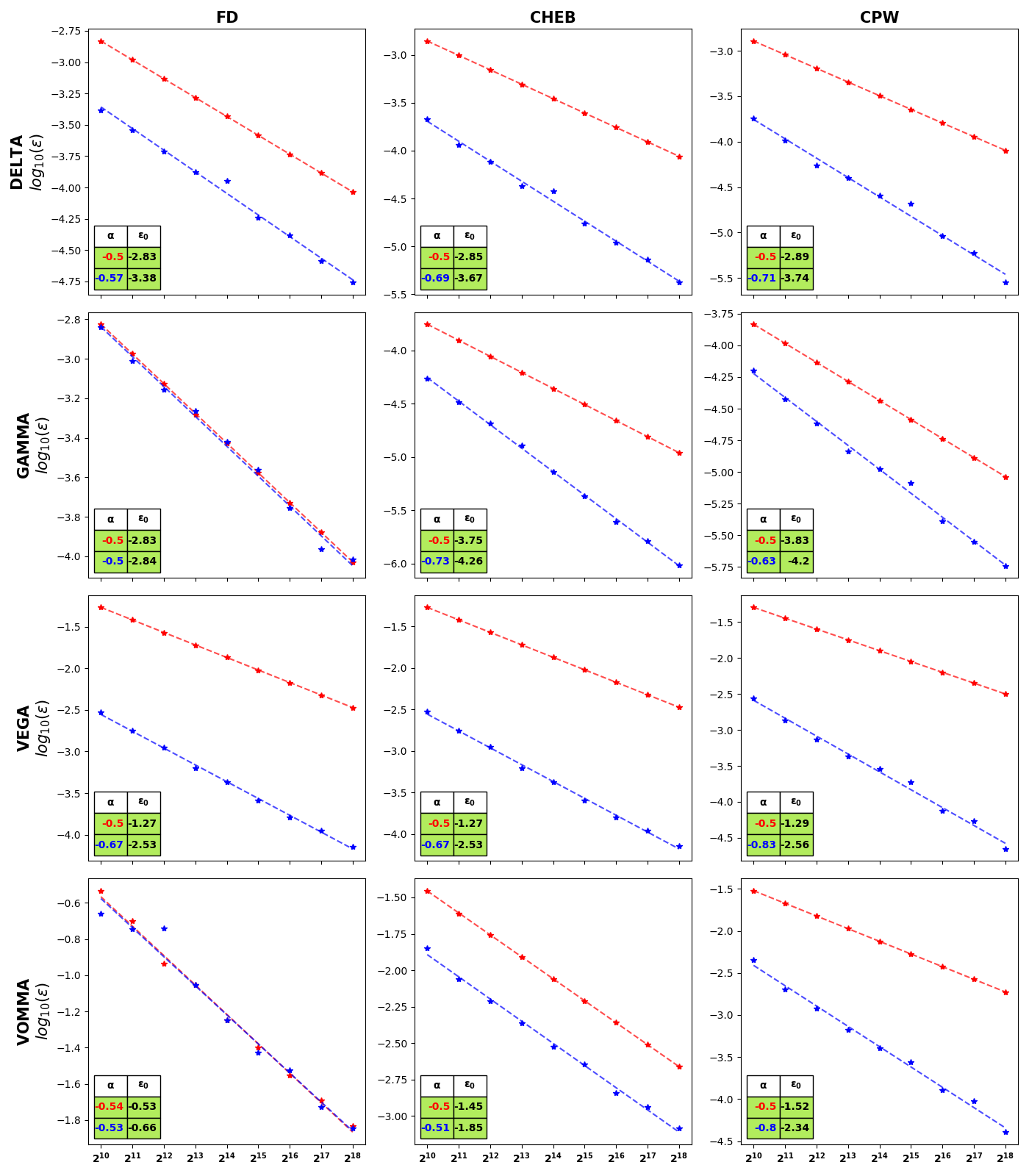}
\caption{Convergence rates for Asian Call option Greeks adopting \textbf{MC+ED} (red) and \textbf{RQMC+BBD} (blue).}
\label{fig:conv_Asian_Greeks}
\end{figure}

\begin{figure}[H]
\centering
\includegraphics[scale=0.4]{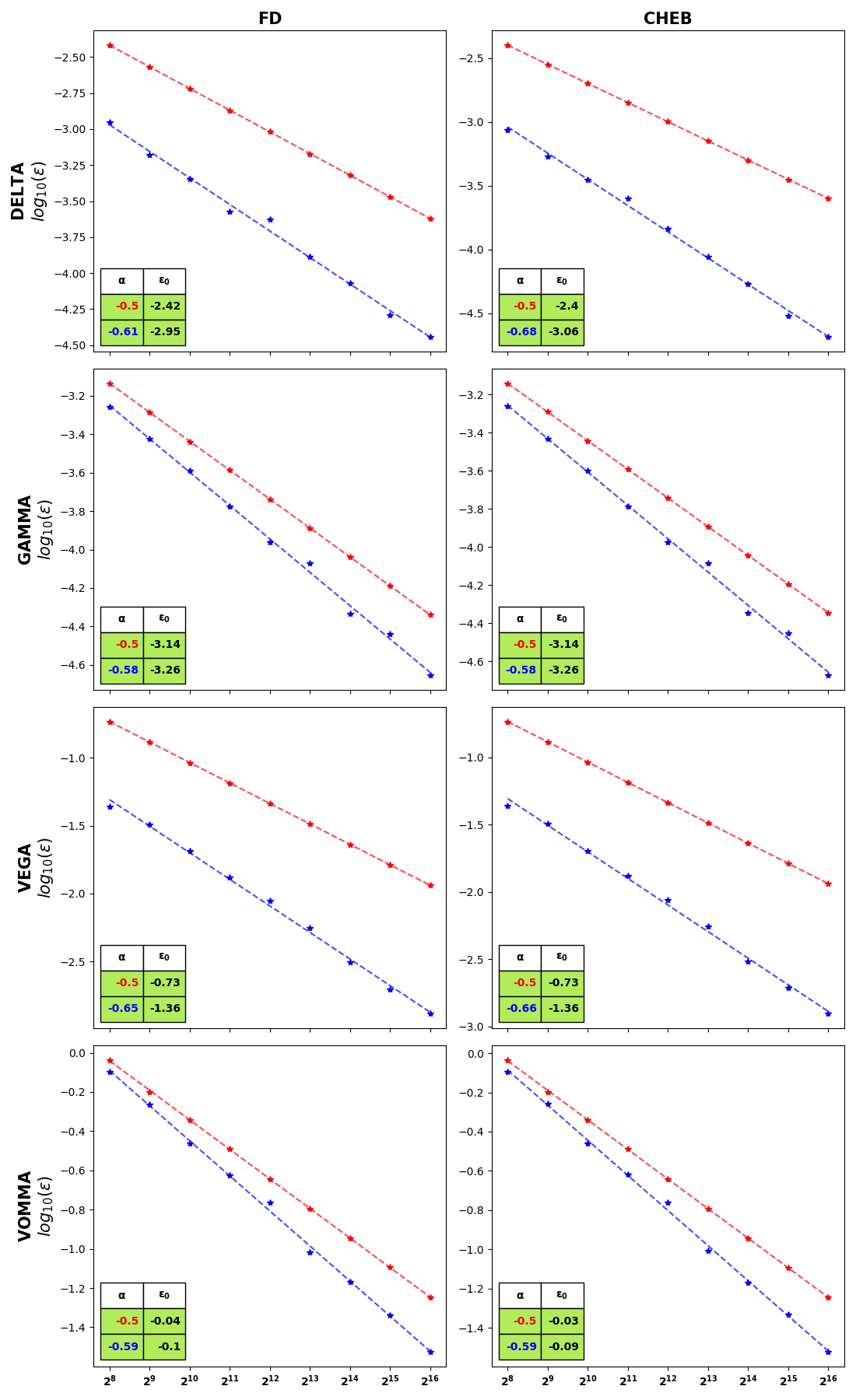}
\caption{Convergence rates for Down-Out Call option Greeks with barrier level at 90 and strike at 100 adopting \textbf{MC+ED} (red) and \textbf{RQMC+BBD} (blue) combined with importance sampling.}
\label{fig:vomma_conv_Asian}
\end{figure}

\section{Conclusions}
In this study, we compared three methods for estimating Greeks: finite difference (FD), Chebyshev interpolation (CI), and conditional pathwise (CPW) with two simulation and path discretization approaches: Monte Carlo with Euler discretization (MC+ED) and randomized quasi-Monte Carlo with the Brownian Bridge discretization (RQMC+BBD). Our analysis focused on both first- and second-order derivatives for two widely used path-dependent options: the discrete barrier option and the arithmetic average Asian option. We also incorporated importance sampling to handle challenging parameter sets.

The primary goal was to evaluate the potential for effective dimension reduction provided by CI and CPW within the framework of global sensitivity analysis (GSA) and to determine the most efficient methods for computing Greeks in terms of accuracy, convergence rate, and computational cost. While the advantages of RQMC over standard Monte Carlo in financial applications are well documented, this study is, to the best of our knowledge, the first to investigate the combination of CI and CPW with RQMC from a GSA perspective.

Our results demonstrate a significant advantage of RQMC+BBD+CI and RQMC+BBD+CPW over the traditional MC+ED+FD approach. The reduced effective dimension highlighted by GSA supports the efficiency of these methods. Revisiting the objectives outlined in the Introduction, we can draw several key conclusions from these findings:

\begin{itemize} 
\item [1)] Pairing CI with RQMC and BBD not only enhances accuracy and smoothness but also leads to faster convergence in the majority of the analyzed cases;

\item [2)] In all scenarios, a significant effective dimension reduction either results in a substantial initial error decrease, an improved convergence rate, or both. Notably, both CI and CPW achieve a lower effective dimension compared to FD, showing greater efficiency for the first-order Greeks and exceptional results for the second-order Greeks when combined with BBD. \end{itemize}

Our numerical experiments indicate that selecting the appropriate methodology, guided by GSA, can drastically reduce the number of simulations required to achieve the desired Greeks' accuracy. While CPW may be impractical in certain situations, CI offers flexibility, and the increased computational cost due to the higher number of nodes, compared to FD, is completely offset by the improved accuracy, particularly when combined with RQMC integration and BBD.

\newpage

\appendix
\section{Conditional pathwise estimates}
Here, we provide a brief summary of the conditional estimates for Down-and-Out Call and Asian Call options, using the notation introduced in Section 3. In what follows, we denote by $n(\cdot)$ the standard normal probability density function and by $N(\cdot)$ the standard normal cumulative probability function.
\subsection{Down-Out Call option}
In the case of the barrier down-out call option, the separating variable condition holds if we take $\psi_d$ and $\psi_u$ as follows:
\begin{equation}
    \begin{split}
        \psi_1 &= \frac{\ln{(K/\Tilde{S}_d)}-\mu t_1}{\sigma \sqrt{t_1}},\quad\quad
        \psi_2 = \frac{\ln{(B/\Tilde{S}_{min})}-\mu t_1}{\sigma \sqrt{t_1}},\\
        \psi_d &= \max{(\psi_1,\psi_2)},\quad\quad
        \psi_u = +\infty, \quad\quad \psi_d^* = \psi_d - \sigma \sqrt{t_1}.
    \end{split}
\end{equation}
Here we used the following notations
\begin{equation}
\begin{split}
    \Tilde{S}_d &= S_0 \exp{\big(\mu(T-t_1) + \sigma (W_T - W_1)\big)},\\
    S_{min} &=\min\limits_{t\in\mathcal{T}} S_t, \quad\quad 
    \Tilde{S}_{min} =\min\limits_{t\in\mathcal{T}} \Tilde{S}_t\\
\end{split}
\end{equation}
and with $\mathcal{T}$ denoting the time grid of barrier observation dates.
Then, the conditional estimate (\ref{conditioning_step}) takes the form:
\begin{equation}
\label{do_call_CPW_estimate}
    G(\mathbf{X}_{-i}, \theta) = e^{r(t_1-T)} \Tilde{S}_d(1 - N(\psi_d^*)) - e^{-rT} K (1 - N(\psi_d)).
\end{equation}
Conditional estimates for the Greeks are then obtained by differentiating (\ref{do_call_CPW_estimate}) w.r.t. desired parameters:
\begin{equation}
    \begin{split}
        \frac{\partial G}{\partial S_0} &= e^{r(t_1 - T)}\frac{\Tilde{S}_d}{S_0}\bigg[N(- \psi_d^*) + \frac{n(\psi_d^*)}{\sigma \sqrt{t_1}} \bigg] -e^{-rT}\frac{K n(\psi_d)}{S_0 \sigma \sqrt{t_1}}, \\
        \frac{\partial^2 G}{\partial S_0^2} &= \frac{1}{S_0^2 \sigma^2 t_1}\bigg(e^{r(t_1 - T)}\Tilde{S}_d n(\psi_d^*) \psi_d - e^{-rT} K n(\psi_d)\psi_d^*\bigg),\\
        \frac{\partial G}{\partial \sigma} &= e^{r(t_1 - T)} \Tilde{S}_d \big[  \Tilde{X}_d N(-\psi_d^*) + n(\psi_d^*)\sqrt{t_1}\big] + \frac{\partial \psi_2}{ \partial \sigma} \big[e^{-rT} K n(\psi_2) - e^{r(t_1 - T)} \Tilde{S}_d n(\psi_2^*) \big]\mathbbm{1}_{\{\psi_2 > \psi_1\}},\\
        \frac{\partial^2 G}{\partial \sigma^2}&= \frac{\partial}{\partial \sigma}\big[ A + B\mathbbm{1}_{\{\psi_2 > \psi_1\}}\big],\\
        \frac{\partial A}{\partial \sigma} & = e^{r(t_1 - T)} \Tilde{S}_d\bigg[
        \Tilde{X}_d
        \big[\Tilde{X}_d N(-\psi_d^*) + n(\psi_d^*)\sqrt{t_1}\big]+ (t_1-T)N(-\psi_d^*)
        - n(\psi_d^*) \frac{\partial \psi_d^*}{\partial \sigma}\big(\Tilde{X}_d + \psi_d^* \sqrt{t_1}\big)\bigg],\\
        \frac{\partial B}{\partial \sigma} &=\frac{\partial^2 \psi_2}{\partial \sigma^2}\big[e^{-rT} K n(\psi_2) - e^{r(t_1 - T)} \Tilde{S}_d n(\psi_2^*) \big],\\
        &+ \frac{\partial \psi_2}{\partial \sigma}\big[-e^{-rT} K n(\psi_2) \psi_2 \frac{\partial \psi_2}{\partial \sigma} - e^{r(t_1 - T)} \Tilde{S}_d n(\psi_2^*) \big( \Tilde{X}_d - \psi_2^*\frac{\partial \psi_2^*}{\partial \sigma} \big) \big].
\end{split}
\end{equation}
\subsection{Arithmetic average Asian Call option}
In the case of the arithmetic average Asian Call option, the separating variable condition holds if we take $\psi_d$ and $\psi_u$ as follows:

\begin{equation}
\psi_d = \frac{\ln{(K/\Tilde{S}_A)}-\mu t_1}{\sigma \sqrt{t_1}},\quad\quad \psi_u = +\infty,\quad\quad \psi_d^*= \psi_d - \sigma \sqrt{t_1} \quad\quad \Tilde{S}_A = \frac{1}{D} \sum\limits_{j=1}^D \Tilde{S}_j.
\end{equation}
In this case, the conditional estimate (\ref{conditioning_step}) takes the form:
\begin{equation}
\label{asian_call_CPW_estimate}
    G(\mathbf{X}_{-i}, \theta) = e^{r(t_1-T)} \Tilde{S}_A (1 - N(\psi_d^*)) - e^{-rT} K (1 - N(\psi_d))
\end{equation}
Conditional estimates for the Greeks are then obtained by differentiating (\ref{asian_call_CPW_estimate}) w.r.t. desired parameters:
\begin{equation}
    \begin{split}
        \frac{\partial G}{\partial S_0} &= e^{r(t_1 - T)}\frac{\Tilde{S}_A}{S_0}\big[ 1 - N(\psi_d^*) \big],\\
        \frac{\partial^2 G}{\partial S_0^2} &= \frac{K e^{-rT}}{S_0^2 \sigma t_1} n(\psi_d),\\
        \frac{\partial G}{\partial \sigma} &= e^{r(t_1 - T)} \big[1- N(\psi_d^*)\big]\frac{\partial \Tilde{S}_A}{\partial \sigma}  + e^{-rT} K n(\psi_d) \sqrt{t_1},\\
        \frac{\partial^2 G}{\partial \sigma^2}&= e^{r(t_1 - T)} \bigg[ -n(\psi_d^*) \frac{\partial \psi_d^*}{\partial \sigma} \frac{\partial S_A}{\partial \sigma} + [1- N(\psi_d^*)]  \frac{\partial^2 S_A}{\partial \sigma^2}\bigg] - e^{-rT} \psi_d K n(\psi_d) \frac{\partial \psi_d}{\partial \sigma} \sqrt{t_1}.
    \end{split}
\end{equation}

\section{Greeks approximation parameters}
\begin{figure}[H]
\centering
\includegraphics[scale=0.5]{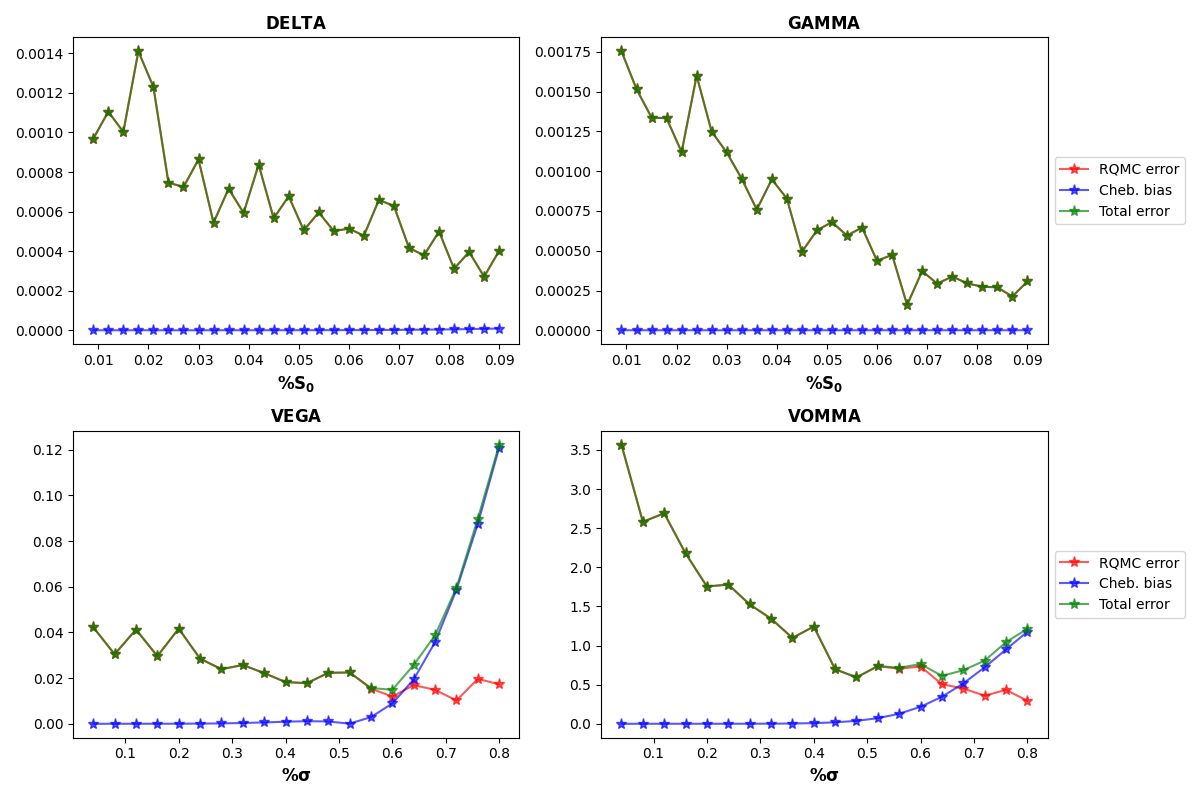}
\caption{Chebyshev error comparison using different interpolation intervals with $2^{16}=65536$ simulations. }
\label{fig:best_cheb_range_cont_do_call}
\end{figure}

When applying FD or CI, choosing the proper parameters to adopt in the approximation method is necessary. In the case of Greeks, an additional error due to numerical differentiation arises (\textit{differentiation bias}) and contributes, along with the MC/RQMC integration error (\textit{in-sample error}), to the sensitivity precision. In this section, we aim to analyze and balance these two errors. While the bias from differentiation cannot be eliminated, the MC/RQMC error decreases as the number of simulations increases. Given the faster convergence seen with RQMC, it becomes more likely to reach a level of precision where the bias has a greater impact. Therefore, we opted for RQMC integration in this analysis. A detailed discussion of the bias-variance problem is described in \cite{Gla03}, Chapter 7.
Generally, increasing the number of points in the grid reduces the derivative approximation bias without affecting the variance of the MC or RQMC estimation. However, adding more points also increases the computational time required for the estimation. On the other hand, using a smaller grid of points typically leads to higher MC or RQMC estimation variance. In the case of FD, selecting the optimal shift is crucial, while CI requires choosing the appropriate width for the interpolation interval.
In this work, we choose to assess separately bias and variance considering a range of different parameters for both FD and CI.
The FD shift parameter had minimal impact on the bias within the range of values examined, and it was selected primarily to limit the MC/RQMC variance. Our focus is on the analysis performed for the CI domain, serving as an example of the methodology used.
To determine an optimal CI interval width $\delta \theta$ (so that the interval considered becomes $[\theta-\delta \theta,\theta + \delta \theta]$), we ran an empirical investigation. 
Two different kinds of errors were compared and weighed: 
\begin{itemize}
    \item the Chebyshev approximation error, computed by using the Chebyshev method on a grid of prices computed using analytical formulas;
    \item the error due to MC or RQMC estimation of the Greeks.
\end{itemize} 
Figure \ref{fig:best_cheb_range_cont_do_call} helps to empirically fix the width that balances the bias $\varepsilon_{bias}$ and the variance $\sigma^2_{RQMC}$ in the simulated results. The total error is computed as
\begin{equation}
    \varepsilon_{tot} = \sqrt{\varepsilon_{bias}^2 + \sigma^2_{RQMC}}.
\end{equation}
We notice, that RMSE and the total errors overlap when the CI bias is much lower than the RMSE error.
Bias estimation can be performed only when an analytical expression for the Greek is available. The results for both the continuous Down-and-Out Call Option and the geometric average Asian Call Option can be found in \cite{Wil013}. We assume that the bias-variance balance holds similarly in the case of the discretely monitored barrier option and arithmetic average Asian option. Observing Figure \ref{fig:best_cheb_range_cont_do_call}, we notice that the RQMC error decreases as the interpolation width increases. On the other hand for Vomma and Vega, the bias introduced by CI is bigger than the RQMC error for large interpolation domains. Given these considerations, the interpolation intervals were chosen to prevent the bias from prevailing over the RQMC error.
\begin{table}[H]
\centering
\begin{tabular}{| c | c | c | c | c |}
\hline
 & \multicolumn{2}{c|}{\textbf{Down-Out Call}}  & \multicolumn{2}{c|}{\textbf{Asian Call}} \\\cline{2-5}

\textbf{Greek} & \textbf{Cheb. Width} & \textbf{FD shift} & \textbf{Cheb. Width} & \textbf{FD shift}\\
\hline
Delta & 5\% of $S_0$ & 0.1\% of $S_0$ & 10\% of $S_0$ & 0.1 \\
\hline
Gamma& 7.5\% of $S_0$ & 0.1\% of $S_0$ & 10\% of $S_0$ & 0.1 \\
\hline
Vega& 40\% of $\sigma$ & 0.03 \% of $\sigma$ & 40\% of $S_0$ & 0.03 \% of $\sigma$ \\
\hline
Vomma& 40\% of $\sigma$ & 0.03 \% of $\sigma$ & 40\% of $S_0$ & 0.03 \% of $\sigma$\\
\hline
\end{tabular}
\caption{Widths for the CI domain for all the computed Greeks of continuously monitored barrier option.}
\label{cheb_widths}
\end{table}

\section{Importance sampling}

A commonly used technique for Barrier Options is to apply importance sampling (IS) as a variance reduction strategy (see, for example, \cite{Gla03}, Section 4.6, and \cite{Jac01}, Section 11.4). In the case of Down-and-Out Call barrier options, IS adjusts the generated pseudo-random or low-discrepancy sequence to avoid the underlying trajectory $S_t$ touching the barrier, while weighting the simulations based on their likelihood. This method focuses computational effort on simulating trajectories that finish in the money, excluding those that drop below the barrier and result in a zero payoff. Moreover, preventing the underlying asset from crossing the barrier eliminates singularities in the sampled trajectory space, thereby smoothing the payoff function. However, IS is inherently dependent on both the payoff and the specific underlying model. For the Down-and-Out Call option, the algorithm (\ref{Euler_BS}) can be modified as follows (see \cite{Gla03}):

\begin{equation}
\label{IS_DOC}
    S_{j+1} = S_j \exp\big[(r -\frac{\sigma^2}{2}) \Delta t_j + \sigma \sqrt{\Delta t_j} N^{-1}(\mathbf{U}) \big]
\end{equation}
with $\mathbf{U}$ defined as:
\begin{equation}
    \mathbf{U}=(1-p(S_j))+\Tilde{\mathbf{U}}p(S_j).
\end{equation}
Here $\Tilde{\mathbf{U}}$ is uniformly distributed and:
\begin{equation}
\begin{split}
    p(S_j)=\mathbb{P}(S_{j+1}\geq B_j | S_j)=N\bigg( \frac{\ln{(S_j/B_j) + (r - \frac{\sigma^2}{2}) \Delta t_j}}{\sqrt{\Delta t_j } \sigma} \bigg).
\end{split}
\end{equation}
Consider $B_j = B$ for $j<D$ and $B_j = K$ for $j=D$, since at maturity it is desired to sample only options in the money.
For each simulated underlying trajectory $S^k=(S_0^k,...,S_D^k)$ , it is important to store the likelihood factor:
\begin{equation}
    L_k=\prod\limits_{j=0}^{D-1} p(S^k_j).
\end{equation}
The payoff outcome resulting from discretization (\ref{IS_DOC}) is weighted considering the above factor:
\begin{equation}
    V=\mathbb{E}^{\mathbb{Q}}[f]\simeq \frac{1}{N} \sum\limits_{k=1}^N f(\mathbf{u}_k)L_k.
\end{equation}

\section{Price results}
Table \ref{tab:prices} presents for completeness results for prices of the analyzed options.

\begin{table}[H]
\centering
\begin{tabular}{| c | c | c | c | c | c | c | c | c |}
\cline{1-9}
  & \multicolumn{4}{c |}{\textbf{ED + MC} $(\alpha = 0.5)$}  & \multicolumn{4}{c |}{\textbf{BBD + RQMC} $(\alpha = 0.74)$} \\\cline{2-9}
 \textbf{Paths} & \multicolumn{2}{c|}{\textbf{Down-Out Call}}  & \multicolumn{2}{c|}{\textbf{Asian Call}} & \multicolumn{2}{c|}{\textbf{Down-Out Call}}  & \multicolumn{2}{c|}{\textbf{Asian Call}} \\\cline{2-9}
 & \textbf{Price} & \textbf{Error} & \textbf{Price} & \textbf{Error} & \textbf{Price} & \textbf{Error} & \textbf{Price} & \textbf{Error}\\
\hline
$2^{10}$ & $5.96901$ & $2.71 * 10^{-2}$ & $3.73727$ & $1.51 * 10^{-2}$ & $5.98531$ & $3 * 10^{-3}$ & $3.71045$ & $7.33 * 10^{-4}$ \\\hline 
$2^{11}$ & $5.95949$ & $1.93 * 10^{-2}$ & $3.69438$ & $1.09 * 10^{-2}$ & $5.98625$ & $2.3 * 10^{-3}$ & $3.71186$ & $4.41 * 10^{-4}$ \\\hline
$2^{12}$ & $5.98656$ & $1.36 * 10^{-2}$ & $3.69959$ & $7.67 * 10^{-3}$ & $5.98820$ & $1.66 * 10^{-3}$ & $3.71174$ & $2.49 * 10^{-4}$ \\\hline
$2^{13}$ & $5.97346$ & $9.6 * 10^{-3}$ & $3.71412$ & $5.42 * 10^{-3}$ & $5.98654$ & $1.01 * 10^{-3}$ & $3.71198$ & $1.37 * 10^{-4}$ \\\hline
$2^{14}$ & $5.98509$ & $6.71 * 10^{-3}$ & $3.71865$ & $3.89 * 10^{-3}$ & $5.98655$ & $7.86 * 10^{-4}$ & $3.71194$ & $9.7 * 10^{-5}$ \\\hline
$2^{15}$ & $5.98832$ & $4.81 * 10^{-3}$ & $3.71473$ & $2.7 * 10^{-3}$ &  $5.98632$ & $5.46 * 10^{-4}$ & $3.71198$ & $5.19 * 10^{-5}$ \\\hline
$2^{16}$ & $5.98632$ & $3.39 * 10^{-3}$ & $3.71244$ & $1.94 * 10^{-3}$ & $5.98604$ & $3.72 * 10^{-4}$ & $3.71193$ & $3.4 * 10^{-5}$ \\\hline
$2^{17}$ & $5.98582$ & $2.4 * 10^{-3}$ & $3.71170$ & $1.37 * 10^{-3}$ & $5.98590$ & $2.53 * 10^{-4}$ & $3.71198$ & $1.99 * 10^{-5}$ \\\hline
$2^{18}$ & $5.98819$ & $1.69 * 10^{-3}$ & $3.71371$ & $9.67 * 10^{-4}$ & $5.98628$ & $1.71 * 10^{-4}$ & $3.71198$ & $1.27 * 10^{-5}$ \\\hline
\end{tabular}
\caption{Prices and integration errors for the Asian Call option ($K = 100$) and the Down-Out Call option ($K = 100, \quad B = 90$).}
\label{tab:prices}
\end{table}

\section*{Disclaimer}
The views and opinions expressed in this paper are those of the authors and do not represent the views and opinions of their employers. They are not responsible for any use that may be made of these contents.

\newpage
\bibliography{biblio}

\newcommand{\etalchar}[1]{$^{#1}$}
\begin{thebibliography}{{BRO}24}

\bibitem[AK21]{Ata21}
E.~Atanassov and S.~Kucherenko.
\newblock Implementation of owen’s scrambling with additional permutations for sobol’ sequences.
\newblock {\em BRODA Ltd., UK}, 2021.

\bibitem[BKS15]{Bia15}
M.~Bianchetti, S.~Kucherenko, and S.~Scoleri.
\newblock Pricing and risk management with high-dimensional quasi monte carlo and global sensitivity analysis.
\newblock {\em Wilmott}, 78:46--70, 2015.
\newblock \href {https://doi.org/10.1002/wilm.10434} {\path{doi:10.1002/wilm.10434}}.

\bibitem[BKW22]{Bil22}
P.~Bilokon, S.~Kucherenko, and C.~Williams.
\newblock Quasi-monte carlo methods for calculations derivatives sensitivities on the gpu.
\newblock {\em SSRN}, 2022.
\newblock \href {https://doi.org/10.2139/ssrn.4227386} {\path{doi:10.2139/ssrn.4227386}}.

\bibitem[{BRO}24]{BRODA}
{BRODA Ltd}.
\newblock High-dimensional {S}obol' sequence generators.
\newblock {\em http://www.broda.co.uk/}, 2024.

\bibitem[BT04]{Tre04}
J.P. Berrut and L.N Trefethen.
\newblock Barycentric lagrange interpolation.
\newblock {\em SIAM Journal}, 46(3):501–517, 2004.
\newblock \href {https://doi.org/10.1137/S0036144502417715} {\path{doi:10.1137/S0036144502417715}}.

\bibitem[CMO97]{Caf97}
R.~E. Caflish, W.~Morokoff, and A.~Owen.
\newblock Valuation of mortgage-backed securities using brownian bridges to reduce effective dimension.
\newblock {\em Journal of Computational Finance}, 10(1):27--46, 1997.
\newblock \href {https://doi.org/10.21314/JCF.1997.005} {\path{doi:10.21314/JCF.1997.005}}.

\bibitem[Gla03]{Gla03}
P.~Glasserman.
\newblock {\em Monte Carlo Methods in Financial Engineering}.
\newblock Springer, 2003.

\bibitem[GSG07]{Gir07}
Ö~Giray, E.~Salta, and A.~Göncü.
\newblock On pricing discrete barrier options using conditional expectation and importance sampling monte carlo.
\newblock {\em Mathematical and Computer Modelling}, 47:484--494, 2007.
\newblock \href {https://doi.org/10.1016/j.mcm.2007.05.001} {\path{doi:10.1016/j.mcm.2007.05.001}}.

\bibitem[Jac01]{Jac01}
P.~Jackel.
\newblock {\em Monte Carlo Methods in Finance}.
\newblock Wiley, 2001.

\bibitem[KFSM11]{Kuc11}
S.~Kucherenko, B.~Feil, N.~Shah, and W.~Mauntz.
\newblock The identification of model effective dimensions using global sensitivity analysis.
\newblock {\em Reliability Engineering and System Safety}, 96(4):440–449, 2011.
\newblock \href {https://doi.org/10.1016/j.ress.2010.11.003} {\path{doi:10.1016/j.ress.2010.11.003}}.

\bibitem[KH22]{Hok22}
S.~Kucherenko and J.~Hok.
\newblock The importance of being scrambled: Supercharged quasi monte carlo.
\newblock {\em Journal of Risk}, 26(1):1–20, 2022.
\newblock \href {https://doi.org/10.21314/JOR.2023.008} {\path{doi:10.21314/JOR.2023.008}}.

\bibitem[KS07]{Kuc07}
S.~Kucherenko and N.~Shah.
\newblock The importance of being global. application of global sensitivity analysis in monte carlo option pricing.
\newblock {\em Wilmott}, 4:2--10, 2007.

\bibitem[MPS21]{Mar21}
A.~Maran, A.~Pallavicini, and S.~Scoleri.
\newblock Chebyshev greeks smoothing gamma without bias.
\newblock {\em Risk.net}, 2021.
\newblock \href {https://doi.org/10.48550/arXiv.2106.12431} {\path{doi:10.48550/arXiv.2106.12431}}.

\bibitem[Owe97]{Owe97}
A~Owen.
\newblock Scrambled net variance for integrals of smooth functions.
\newblock {\em The Annals of Statistics}, 25(4):1541–1562, 1997.
\newblock \href {https://doi.org/10.1214/aos/1031594731} {\path{doi:10.1214/aos/1031594731}}.

\bibitem[Owe03]{Owe03}
A.~Owen.
\newblock The dimension distribution and quadrature test functions.
\newblock {\em Stat Sinica}, 13(1):1–17, 2003.

\bibitem[RBS20]{Ren20}
S.~Renzitti, P.~Bastani, and S.~Sivorot.
\newblock Accelerating cva and cva sensitivities using quasi-monte carlo methods.
\newblock {\em Wilmott}, 108:78--93, 2020.

\bibitem[SAKK11]{Sob11}
Ilya~M Sobol', Danil Asotsky, Alexander Kreinin, and Sergei Kucherenko.
\newblock Construction and comparison of high-dimensional {S}obol' generators.
\newblock {\em Wilmott}, 2011(56):64--79, 2011.

\bibitem[SBK21]{Sco21}
S.~Scoleri, M.~Bianchetti, and S~Kucherenko.
\newblock Pricing and risk management with high-dimensional quasi monte carlo and global sensitivity analysis.
\newblock {\em Wilmott}, 116:66--83, 2021.

\bibitem[Sob93]{Sob93}
I.M. Sobol.
\newblock Sensitivity estimates for nonlinear mathematical models.
\newblock {\em Mathematical Modeling and Computational Experiments}, 4:407--414, 1993.

\bibitem[Sob01]{Sob01}
I.M. Sobol’.
\newblock Global sensitivity indices for nonlinear mathematical models and their monte carlo estimates.
\newblock {\em Mathematics and Computers in Simulation}, 55(1-3), 2001.

\bibitem[STS{\etalchar{+}}07]{Mau07}
I.M. Sobol, Tarantola, D.~S., Gatelli, S.~Kucherenko, and W.~Mauntz.
\newblock Estimating the approximation error when ﬁxing unessential factors in global sensitivity analysis.
\newblock {\em Reliability Engineering and System Safety}, 92(7):957–60, 2007.
\newblock \href {https://doi.org/10.1016/j.ress.2006.07.001} {\path{doi:10.1016/j.ress.2006.07.001}}.

\bibitem[Wel97]{Wel97}
B.~Welfert.
\newblock Generation of pseudospectral differentiation matrices.
\newblock {\em SIAM Journal on Numerical Analysis}, 34(4):1640–1657, 1997.
\newblock \href {https://doi.org/10.1137/S0036142993295545} {\path{doi:10.1137/S0036142993295545}}.

\bibitem[Wil13]{Wil013}
Paul Wilmott.
\newblock {\em Paul Wilmott on quantitative finance}.
\newblock John Wiley \& Sons, 2013.

\bibitem[ZW20]{Zha20}
C.~Zhang and X.~Wang.
\newblock Quasi-monte carlo-based conditional pathwise method for option greeks.
\newblock {\em Quantitative Finance}, 20(1):1--19, 2020.

\end{thebibliography}
\end{document}